\documentclass[aps,prl,twocolumn,groupedaddress]{revtex4}
\newtheorem{thm}{Theorem}

\newtheorem{defin}{Definition}

\newtheorem{cor}{Corollary}
\newtheorem{claim}{Claim}
\newtheorem{lem}{Lemma}

\newtheorem{rem}{Remark}

\begin{document}


\title{Weak realism, counterfactuals, and decay of geometry at small scales}


\author{Charles Tresser}
\email[]{charlestresser@yahoo.com}
\affiliation{IBM, P.O. Box 218, Yorktown Heights, NY 10598, U.S.A.}


\date{\today}

\begin{abstract}
Two typical entanglements will be shown to stand on opposite sides on the issue of \emph{instrumental realism}, the issue of whether (as for EPR in the original form or EPRB, Bohm's versions using spin) or not (as for GHZ) observables have values that preexist measurement.  Instrumental realism in the EPR context helps us prove that in some special circumstances, one can get simultaneous knowledge of two conjugate quantities, which in particular make sense together. This shatters the axiomatic presentation of Quantum Mechanics.  This simultaneous knowledge of two conjugate quantities is an elaboration on 1935 work by Schr{\"o}dinger, hence the name, \emph{Schr{\"o}dinger Unorthodoxy Theorem}, given to the second main result that is obtained with little effort from the result on instrumental realism. Once the axiomatic edifice of Quantum Mechanics is broken, we can let go the completeness of the wave function as suggested in EPR, and like Einstein ``absolutely hold fast" on locality.  The EPR paper of mid-1935 gets here contrasted, from a new point of view, with a 1936 text by Einstein where Einstein avoids using counterfactuals. Counterfactuals get a precise definition and the corresponding concept is used all along, but may be new under an old name.  We provide a short critical review of Bell's 1964 paper.  Then, a small modification of arguments for the Schr{\"o}dinger Unorthodoxy Theorem will let appear a simple conservation law, combined with the Malus Law,  as the origin of the correlation in Bell's version of EPRB: this is our third main result.  As the fourth main result, a last use of the concept of counterfactual yields the decay of geometry at small enough scale.  This opens a new world of interpretation of aspects of Quantum Mechanics, aspects that range from measurement and the need of classical physics to views on what realism should mean in microphysics.
\end{abstract}

\maketitle

%
%
%
%
%
\section{1: Introduction}\label{sec:introduction}
The word \emph{entanglement}, but not the concept, was apparently first used in the context of 
Quantum Mehanics (or \emph{QM}) by Schr{\"o}dinger in \cite{Schrodinger1935a}, arguably the first significant article about the Einstein, Podolsky, Rosen (or \emph{EPR}) paper \cite{EPR}, written to continue in the line of thought of \cite{EPR} rather than attack it.  It is well documented (see \cite{Jammer1974}, \cite{FineShaky}) that just after the publication of \cite{EPR}, Schr{\"o}dinger shared with Einstein his amusement at watching the disarray of the other QM gurus confronted with that paper, and got in exchange Einstein's private own view on the issues covered by \cite{EPR}, a view somewhat different from what Podolsky had written as the acting author of that historical paper.  After expounding the central role of entanglements in QM, Schr{\"o}dinger mentions in \cite{Schrodinger1935a}, the fact that \emph{two conjugate variables can be known on a single particle} but in a rather cryptic way (see also accounts on this paper in \cite{Jammer1974} and \cite{Fine2}). 

\medskip
In \cite{Schilpp} pp. 681-682, about the dispute with the QM orthodox views about EPR, and after telling us that Bohr had the best perspective among the orthodoxes, Einstein (who began using the word ``paradox" in contexts like in \cite{EPR} years before the published version, and used \emph{$\psi$-function} for the \emph{wave-function}) writes (throughout the paper, I use square brackets to provide context that help make sense of the cited texts and I always reproduce italics and quotation marks exactly as in the quoted source when they are on parts of the quotation) :

\smallskip
\noindent
``Translated into my own way of putting it, he [Niels Bohr] argues as follows:

\smallskip
If the partial systems $A$ and $B$ form a total system which is
described by its $\psi$-function $\psi (AB)$, there is no reason
why any mutually independent existence (state of reality) should
be ascribed to the partial systems $A$ and $B$ viewed separately,
\emph{not even if the partial systems are spatially separated from
each other at the particular time under consideration.} 
The assertion that, in this latter case, the real situation of $B$could not be (directly) influenced by any measurement taken on $A$
is, therefore within the framework of quantum theory, unfounded and (as the paradox shows) unacceptable.

\smallskip
By this way of looking at the matter it becomes evident that the
paradox forces us to relinquish one of the following two
assertions:
\begin{description}
\item[({\bf 1})] the description by means of the $\psi$-function is
\emph{complete}
\item[({\bf 2})] the real states of spatially
separated objects are independent of each other.
\end{description}

In the other hand, it is possible to adhere to ({\bf 2}), if one
regards the $\psi$-function as the description of a (statistical)
ensemble of systems (and therefore relinquish ({\bf 1})). However,
this view blasts the framework of the ``orthodox quantum theory."\,"

\smallskip
\noindent So is the ({\bf 1})\emph{vs}({\bf 2}) alternative set by Einstein.  On this very alternative, Einstein clearly chose ({\bf 2}).  In the same book \cite{Schilpp} as just cited (but in the first volume in the two volumes editions), he wrote p. 85:

\smallskip
\noindent ``But on one supposition we should, in my opinion, absolutely hold fast: the real factual situation of the system $S_2$ is independent of what is done with the system $S_1$, which is spatially separated from the former."

\noindent
Bell accepted this alternative.  But, as clearly as Einstein had chosen to relinquish ({\bf 1}), Bell chose to relinquish ({\bf 2}) \cite{Bell}  (see also \cite{Bell2001}  or \cite{Bell2004} that contain respectively reprints and new prints of most of the papers of Bell that are related to the foundations of QM) while Rosen who chose to relinquish ({\bf 1}) in 1935 \cite{EPR}, chose to relinquish ({\bf 2}) in 1985 in  a paper \cite{Rosen1985} about \cite{EPR}.  The way I understand it, Rosen's late choice and Bell's choice are both due to the same decision to follow what classical QM dictates in the choice  ({\bf 1})\emph{vs}({\bf 2}).  Notice that long range changes in the wave function by virtue of conservation of the norm are not the issue, as they are linked to the obvious lack of Lorentz invariance of non-relativistic QM: it is what would remain a problem in any relativistic QM that is concerned when one speaks about the locality \emph{vs} non-locality issue.  Each of Einstein, Rosen, and Bell understood clearly about relinquishing ({\bf 1}) that, ``this view blasts the framework of the ``orthodox quantum theory."   I believe (and this strongly motivates the present work) that, if a violation of QM as strong as Schr{\"o}dinger's statement in \cite{Schrodinger1935a} that two conjugate variables can be known on a single particle can be further justified, this leaves no serious reason to support ({\bf 2}) over ({\bf 1}).  Unfortunately, recent good sense based arguments for supporting ({\bf 2}) that I know about (see for instance \cite{Omnes} and \cite{Griffiths}) are formulated without even mentioning the  ({\bf 1})\emph{vs}({\bf 2}) alternative.  However, the simplest Bell type theorem presented in \cite{Tresser2005} allows one to get false inequalities as in previously known Bell type theorems, but with no room for any effect of non-locality, all being due to counterfactuals: this provides another strong argument to fully abandon non-locality. 

\medskip
The paper is organised as follows:
\begin{description}
\item[-] A short description of the so-called EPR and \emph{GHZ} (for Greenberger, Horne, Zeilinger: \cite{GHZ1989}) entanglements (the later  in the 3 particles form proposed by Mermin \cite{Mermin1990GHZ3}: see also \cite{GHSZ1990GHZ3}) is given in Section 2. I also discuss there
EPRB, the Bohm version \cite{Bohm}  of  EPR where one considers the spin of spin-$\frac{1}{2}$ particles along two orthogonal axes. 
\item[-]  Discussions  of the method of  \emph{rejection of conclusions from counterfactual experiments} and about \emph{which kind of realism one needs when studying entanglements} are presented in Section 3 where I also report on Einstein's own discussion of the EPR issue, and present remarks on the EPR paper itself and its links to counterfactuals, some of which may be new.
\item[-]  This is more than I need to allow me to restate and further justify in Section 
4 the heretic behavior of EPR particles pointed out by Schr{\"o}dinger in \cite{Schrodinger1935a} (see also \cite{Schrodinger1935b}), in what 
I call the \emph{Schr{\"o}dinger Unorthodoxy Theorem} (or  Claim \ref{claim:C1}). More precisely, after arguing to justify \emph{weak realism} (mostly the fact that \emph{a precise form of realism holds true in the EPR context}), I will show that such realism leads almost immediately to the Schr{\"o}dinger Unorthodoxy Theorem. 
\item[-]  I next briefly discuss in Section
5 Bell's extension to arbitrary angles of EPRB, (and correlatively of the Bohm-Aharonov version \cite{BohmAharonov1957} that deals with the polarization of photons rather than with the spin of spin-$\frac{1}{2}$ particles).  
\item[-]  A small adaptation of the argument for the  Schr{\"o}dinger Unorthodoxy Theorem  then allows me to provide in Section
6 a conservation-law-based justification of the correlations that QM predicts in Bell's extension of EPRB. 
\item[-]  After so fighting aspects of QM orthodoxy, and Bell's ruling clan, I will back-up in Section 
7 one of the most controversial stands of the Copenhagen Interpretation, in fact what Bell may have hated most in the classical views (while he backed strongly the classical stand on the issue raised in \cite{EPR}).  Using the method of rejection of conclusions from counterfactual experiments, I will defend that \emph{the need for classical physics to found quantum physics comes from the very fact that geometry does not make any physical sense at small enough scale}.  This decay of physically meaningful geometry at small enough scale is a trivially established thesis as we will see; however it justifies the very non-trivial fact, based on the Uncertainty Principle, that coordinates do not make classical sense for particles.
\end{description}
%
%
\section{2: Two gedanken experiments that have made it to the lab}{\label{sec:ExperimentalSetting}}
\subsection{2.1: The (gedanken) experiments of the EPR type.}\label{sub:EPR}
The \emph{entanglements} in what I call \emph{the EPR case} in this paper  will be 2-particles states that cannot be written as tensor products. Particles whose states were known before a known interaction and that get separated after the interaction are good examples.  This includes pairs created from energy such as electron-positron pairs that will be our standard examples all along (hence the ${\bf e}$ and ${\bf p}$ symbols).  The entangled state describes the global system after the particles have separated, to become systems $I$  and $II$.  I will let $u_\alpha(x_1)$ (where $x_1$ stands for the the variables used to describe the system $I$) represent the eigenvectors of some observable $A$ for system $I$.  Depending on whether one has a discrete spectrum (as in the case of the spin along some direction $\vec{a}$ as suggested by Bohm who considers spins along two orthogonal directions) or a continuous spectrum (as in the case of momentum following \cite{EPR}), the wave function can be written as follows:
\begin{description}
\item[In the discrete case:]  Following \cite{EPR}, let $a_1,a_2,a_3,\cdots$ be the eigenvalues of some physical quantity $A$ pertaining to system $I$ and $u_1(x_1), u_2(x_1), u_3(x_1),\cdots$ be the corresponding eigenfunctions, with $x_1$ standing for the variable used to describe system $I$. Then Podolsky tells us that $\Psi$,  considered as a function of $x_1$ can be expressed as
\begin{equation}\label{Psi1}
\Psi(x_1,x_2)=\Sigma _{n=1}^\infty \psi _n(x_2)u_n(x_1)\,,
\end{equation}
where $x_2$ stands for the the variables used to describe the system $II$. Here the $ \psi _n(x_2)$'s are to be merely regarded as the coefficients of the expansion of $\Psi$ into a series of orthogonal functions $u_n(x_1)$. 
By \emph{Wave Packet Reduction} (WPR), if the quantity $A$ is measured to $a_k$, then the wave function of system $I$ is given by $u_k(x_1)$ and system $II$ is left in the state with wave function $\psi_k(x_2)$, and Podolsky also notices that the set of functions $u_n(x_1)$ is determined by the choice of the physical quantity $A$ (leaving implicit that the $\psi_n(x_2)$'s also depend on the choice of $A$).

\smallskip
\noindent
However, still following \cite{EPR} (see p. 779):

\smallskip
\noindent
``If, instead of this,  we had chosen another quantity, say $B$" 

\smallskip
\noindent
(which may in particular be an observable conjugate to $A$) having the eigenvalues  $b_1,b_2,b_3, \cdots$ and $v_1(x_1), v_2(x_1), v_3(x_1),\cdots$ as the corresponding eigenfunctions, we would have obtained 
\begin{equation}\label{Psi1'}
\Psi(x_1,x_2)=\Sigma _{n=1}^\infty \phi _m(x_2)v_m(x_1)\,,
\end{equation}
instead of Equation \ref{Psi1} for the $\Psi$ function, with the $\phi_m$'s as the new coefficients. Then, if $B$ is measured to be $b_j$, we conclude from WPR that after the measurement system $I$ is left in the state given by $vi_j(x_1)$ and that system $II$ is left in the state with wave function $\phi_j(x_2)$.

\smallskip
This sets the stage for either accepting an incomplete description by the wave function or admitting the lack of locality, but this is classical material that I will not revisit in details: I need merely the setting since the strategy will be to try improving on \cite{Schrodinger1935a} (which has not been too often tried) rather than having one more frontal attack along the line of \cite{EPR}.

\medskip
In the case of the spin of spin-$\frac{1}{2}$ particles (a configuration often called \emph{EPRB} in the literature), following  Bohm \cite{Bohm} I will consider the so-called \emph{singlet state} (see, \textit{e.g.,} \cite{Bohm}, p. 400)  that is rotation invariant and given by:
\begin{equation}\label{Singlet}
\Psi(x_1,x_2)=\frac{1}{\sqrt{2}}(| +\rangle _1\otimes| -\rangle_2-| -\rangle_1\otimes|
+\rangle_2)\,,
\end{equation}
and there is a corresponding case for photon polarization that is more practical for experiment but that will not be used here \cite{BohmAharonov1957}. The singlet state is particularly simple because, since it gets the same expression in any orthogonal basis, WPR analysis allows one to readily see that the two particles inherit opposite signs. Notice that in this case, the spins along orthogonal axes provide an example of conjugate observable leading to exactly analogous $\Psi$ functions.
 
\item[In the continuous case:] 

For instance (using an example from \cite{EPR}), assume that systems $I$ and $II$ are two particles in one dimensional motion and that:
\begin{equation}\label{PsiCont}
\Psi(x_1,x_2)=\int _{-\infty}^\infty e^{\frac{2\pi  i}{h}(x_1-x_2+x_0)p}dp\,,
\end{equation} 
where $x_0$ is some constant.  Using then the momentum of the first particle as the first observable $A$, the eigenfunctions of $A$ will be
\[
u_p(x_1)=e^{\frac{2\pi  i}{h}px_1}\,.
\] 
Thus, being now in the continuous spectrum case, one gets instead of Equation \ref{Psi1}: 
\begin{equation}\label{Psi2}
\Psi(x_1,x_2)=\int _{-\infty}^\infty \psi _p(x_2)u_p(x_1)dp\,,
\end{equation}
where
$$
\psi _p(x_2)=e^{\frac{-2\pi  i}{h}(x_2-x_0)p}\,,
$$
and we notice that  $\psi _p (x_2)$ is the eigenfunction of the operator
$$
P=({\frac{h}{2\pi  i}}){\frac{\partial x_2}{ \partial}}\,,
$$
corresponding to the eigenvalue $-p$ of the momentum of the second particle.

\smallskip
On the other hand, if now $B$ is taken as the coordinate of the particle that constitutes system $I$, it has for eigenfunction
$$
v_x(x-1)=\delta(x_1-x)\,,
$$
corresponding to the eigenvalue $x$, where we have used $\delta$ for the Dirac delta-function.
Equation \ref{Psi1'} in this case is replaced by
$$
\Psi(x_1,x_2)=\int _{-\infty}^\infty \phi _x(x_2)x_x(x_1)dp\,,
$$
where
$$
\phi _x(x_2)=\int _{-\infty}^\infty e^{\frac{-2\pi  i}{h}(x-x_2+x_0)p}dp=h\delta(x-x_2+x_0)\,.
$$
This $\phi_x$ is now the eigenfunction of the operator $Q=x_2$ corresponding to the eigenvalue $x+x_0$ for the coordinate of the particle that constitutes system $II$.  

At this point of the discussion in \cite{EPR}, Podolsky then uses the fact that $A$ and $B$ are conjugate observables in what seems to me to be a counterfactual-based reasoning, based  on  his use of ``If, instead of this,  we had chosen another quantity, say $B$", locality, and elements of reality (but see Subsection 3.4, and above all, see \cite{EPR} pp.779-780). 

\end{description}
I assume symmetry  (total spin zero, total momentum zero, etc.,) except otherwise stated (as we have done implicitly above, by not imposing $x_0=0$), although symmetry is not necessary but often simplifies the discussion.  Aspect of symmetry may signify constraints that are  (too) hard to enforce, in particular because of the Uncertainty Principle, as would be for instance imposing $x_0=0$ in Equation \ref{PsiCont} above. 

\medskip
As for spin, like Bohm in 1951 \cite{Bohm}, in the EPR context I will first only consider measurements of the singlet state described by Equation \ref{Singlet} along two axis, say $x$ and $z$, orthogonal to each other and to the axis $y$ of propagation. Later, starting in Section 5 (after a brief comment in Section 4, following Remark \ref{rem;Malus}) the ``$x$ and $z$ only" constraint is to be relaxed, following the work of Bell  \cite{Bell}.  In the present theoretical paper, I will mostly ignore the 1957 suggestion 
in \cite{BohmAharonov1957} to replace the spin of spin-$\frac{1}{2}$ particles by photons polarizations as is done in most if not all actual EPR experiments (see for instance \cite{AspectEtAl1982}, \cite{Innsbruck1998} and references therein).

\medskip
\subsection{2.2: The (gedanken) experiments of the GHZ type.}\label{sub:GHZ}
The GHZ setting is a type of entanglement different from what one studies in the EPR context. Initially conceived with 4 particles \cite{GHZ1989},  such entanglements were built to show that the so called \emph{Einstein local realism}, a concept forged to be proved naively wrong, was indeed more obviously wrong (with ```more" in the sense of ``without calling upon statistics") that could be supposedly established by Bell inequalities \cite{Bell}. Further entanglements of the GHZ type were later conceived with 3 particles \cite{Mermin1990GHZ3},   \cite{GHSZ1990GHZ3}. I use the version in \cite{Mermin1990GHZ3} that has made it to the lab and to textbooks (see, \textit{e.g.,}  pp. 152-153 in \cite{Peres1993} or pp. 186-190 in \cite{LeBellac2003}).  Restricting to the spin part, the state that one considers reads:
\begin{widetext}
\begin{equation}\label{Mermin_GHZ1}
\Psi(x_1,x_2,x_3)=\frac{1}{\sqrt{2}}(| +\rangle_1\otimes| +\rangle_2 \otimes| +\rangle_3-| -\rangle_1\otimes|
-\rangle_2 \otimes| -\rangle_3)\,.
\end{equation}
\end{widetext}
The three particles of this GHZ state (\ref{Mermin_GHZ1}) travel out from near $(0,0,0)$ in the plane $y=0$, with the particle labelled $k\in~ \{1,2,3\}$ going approximately along the half-line starting at the origin and making an angle $2 k\pi /3$ with some reference half-line in the plane $y=0$.  For each $k\in \{1,2,3\}$, and $w\in \{x,y,z\}$, $\sigma _w(k)$ is the spin operator along the axis $w(k)$, where:

- $y(k)\equiv y$ is the vertical axis, oriented positively upward,

- $z(k)$ is the axis along which particle $k$ travels, 

- $x(k)$ is orthogonal to $y$ and $z(k)$ and oriented positively counterclockwise. 

We have for any $k\in \{1,2,3\}$:
$$\quad\,\,\sigma_x(k)| +\rangle_k=| -\rangle _k\qquad\,\, \sigma_x(k)| -\rangle_k =| +\rangle _k\,,$$
$$\,\,\,\,\,\,\,\quad\sigma_y(k)| +\rangle _k =i| -\rangle _k\qquad  \sigma_y(k)| -\rangle _k =-i|+\rangle _k\,.$$

\noindent
Then, $\Psi(x_1,x_2,x_3)$ is:

- \textbf{E(+)}) An eigenvector for each of $\sigma _x(1)\sigma _y(2)\sigma _y(3)$,  $\sigma _y(1)\sigma _x(2)\sigma _y(3)$, and  $\sigma _y(1)\sigma _y(2)\sigma _x(3)$ with eigenvalue 1,

- \textbf{E(-)}) An eigenvector for $\sigma _x(1)\sigma _x(2)\sigma _x(3)$ with eigenvalue -1.

\noindent 
From this one can easily deduce two facts formalized in the following easy lemmas:
\begin{lem}\label{lem:noY}
The quantities that are measured by the $\sigma _y(k)$'s cannot be known before the corresponding measurements are performed.
\end{lem}
\textbf{Proof of Lemma \ref{lem:noY}} 
For otherwise, one could predict the values $s_x(k)$ that would be measured by all of the $\sigma _x(k)$'s.
But then, denoting respectively by  $s_y(k)$ the supposedly known values of the measurements  $\sigma _y(k)$,  by \textbf{E(+)}:
 $$
 s_x(1)s _y(2)s _y(3)=s _y(1)s _x(2)s _y(3)=s _y(1)s _y(2)s _x(3)=1\,,
$$
which using: 
$$s _y(1)^2=s _y(2)^2=s _y(3)^2=1\,,$$ 
yields:
\begin{equation}\label{Mermin_GHZ2}
s_x(1)s _x(2)s _x(3)=1\,.
\end{equation}
Since \textbf{E(-)} reads:
\begin{equation}\label{Mermin_GHZ3}
 s_x(1)s _x(2)s _x(3)=-1\,,
\end{equation}
the comparison of Equations (\ref{Mermin_GHZ2}) and (\ref{Mermin_GHZ3}) provides the contradiction that we seek to conclude the proof.  More precisely, this proves that not all the  $\sigma _y(k)$'s can be known before measurement is performed: the rest is by symmetry over the indices, assuming that all the future decisions are equally possible.

$\qquad\qquad\qquad\qquad\qquad\qquad\qquad\qquad\qquad\qquad$\textbf{Q.E.D.}
\begin{lem}\label{lem:noX}
The quantities that are measured by the $\sigma_x(k)$'s cannot be known before the corresponding measurements are performed.
\end{lem}
\textbf{Proof of Lemma \ref{lem:noX}}
For otherwise, denoting the pre-existing values of the measurements $\sigma _x(k)$ respectively by $s_x(k)$, assume that some first measurement $\sigma _y(j)$ is performed. Without loss of generality, we can assume that $j=1$, with a result $s_y(1)$ for the measurement. Then by \textbf{E(+)} we would be able to predict with certainty the values  $s_y(2)$ and  $s_y(3)$ respectively for the measurements $\sigma _y(2)$ and $\sigma _y(3)$, from which the same contradiction as obtained above for the previous lemma follows readily. This proves that not all the  $\sigma _x(k)$'s can be known before measurement is performed: the rest is by symmetry over the indices,  assuming again that all the future decisions are equally possible.

$\qquad\qquad\qquad\qquad\qquad\qquad\qquad\qquad\qquad\qquad$\textbf{Q.E.D.}

\smallskip
These two lemmas are enough to show that realism in the sense of Definition 2 below is not true in general, but I notice that a small modification of the arguments (one just has to consider all possible outcomes) yields the following stronger result:

\begin{thm}\label{lem:noYnoX}
The quantities that are measured by the $\sigma _y(k)$'s and the  $\sigma _x(k)$'s cannot make sense before measurement is performed.
\end{thm}
Like in the case of EPR, experiments have been done on the GHZ entanglement: see for instance \cite{BPWZ1999} and references therein. Some statistical analysis of the GHZ experiments, which take serious account of the less than perfect performance of the captors,  such as \cite{SzaboFine2002}  and \cite{HessPhilipp2004} and the critical papers responding to these attacks on paper that are more in the line of the work of Bell will not be considered here, and neither will other entanglements such as in \cite{Hardy1993}.  

\smallskip
After the above treatment of aspects of GHZ was written, I had the privilege to have a discussion with Itamar Pitowsky.  I then learned from Itamar of the paper \cite{Pitowsky} which discusses the relation between the statements:

\centerline{ \emph{``Every time I measure $A$ I get the result $a$"}}
 
\centerline{and} 
 
\centerline{ \emph{``The value of $A$ is $a$"}.} 
 
\noindent
and which uses the same context of GHZ that I have used here.  The reader will immediately notice that 
\cite{Pitowsky} and the present section are strongly related. There is however enough of difference in the issues and in the conclusions for me to stick to the text above rather than providing a commented reference to that previous work.  Furthermore,  \cite{Pitowsky} claims to  show that:

\centerline{ \emph{``Every time I measure $A$ I get the result $a$"}}
 
\centerline{ \textbf{$\not\Rightarrow$}} 
 
\centerline{ \emph{``The value of $A$ is $a$".}} 
 
\noindent
but the proof, which uses the context of GHZ, does that in a counterfactual way which may be at best irrelevant for our purpose in the present paper.  I must say that the discussion that I had (and during which I was introduced by my interlocutor to further quite fascinating work of his) was closer to my understanding than what I can read from \cite{Pitowsky}.  I may well misunderstand \cite{Pitowsky}, or Itamar Pitowski may have projected upon me his present understanding of a problem that he dealt with 15 years ago.  In particular I understood from Pitowsky that he does not accept now) the sort of action at a distance that is the trademark of the type of strong non-locality that the present paper aims at getting rid of.  To the contrary, one of the first claims of  \cite{Pitowsky} is about establishing such strong non-locality in the GHZ context.  I will need to comment again on \cite{Pitowsky} later on: see Remark \ref {rem:Pitowsky} in Subsection 4.1.
%
%
\section{3: Counterfactuals, some EPR discussions, and a nomenclature for instrumental realism}\label{sec:CounterEinRealism}
\subsection{3.1: Counterfactuals.}\label{sub:Counterfactuals}
\begin{defin}
I loosely define \emph{counterfactual gendanken experiments} (or simply \emph{counterfactuals}) as thought experiments that cannot be performed because performing them would violate Physics (like getting back in time to redo an experiment; a different example will be discussed later).
\end{defin}
Notice that the word ``counterfactual" is used in the literature with a different sense, and it seems in fact with several other senses.  The definition given here is the one that I find most useful.  During a lecture that I had the pleasure and privilege to give at the Technion on January 9, 2005 (a few day after Asher Peres passed away), Petra Scudo and others in the audience resented my use of the word counterfactual.  In fact so evideently that I kept the same definition but created an onomatopoeia beginning by ``counter' and ending in a funny sound in order to go on with my lecture rather than get sidetracked: readers who have strong feelings about what the word ``counterfactual" should mean can thus replace any further occurrence of this word by the word ``blip": I am not yet ready to quit the fight on what the word should mean. The fact that counterfactual experiments  can lead to non-physical conclusions is rather obvious.  That

\smallskip
\noindent 
\emph{``such reasoning are like proofs based on wrong figures in geometry: at best one cannot trust the result"},

\smallskip
\noindent is a remark that was made to me by Ed Spiegel when I told him that I had began to realize that
\emph{gedanken} experiments should be classified according to being doable or not, and further:

- Discriminated according to the theory (or family of theories) to be tested.

- Or considered as \emph{absolute} (or \emph{q-absolute} with ``q"
standing here for \emph{quasi}) meaning that the judgement of
being doable would be made against all accepted theories.

\smallskip
There are many ways in which a \emph{gedanken} experiment can 
be counterfactual, some of which being more prone than others to occur accidentally.  For 
instance playing with \emph{``either this or that"}, one may easily unwillingly generate the counterfactual to Quantum Mechanics consisting in redoing a given experiment, at least in principle.

\subsection{3.2: Einstein's own discussion of the EPR setting}\label{sub:Einstein'sProof}
The declared intent of the EPR paper \cite{EPR} was to prove
\emph{the incompleteness of Quantum Mechanics} in the following
very precise sense (whatever bigger goal each of the authors had,
if any):

\smallskip
\noindent (*) \textit{the quantum-mechanical description of
physical reality given by  wave functions is not complete.}

\smallskip
\noindent Here, the words \emph{``physical reality"} should be taken in their usual non-technical sense, 
despite the controversial definition in \cite{EPR} of  \emph{elements of physical reality}, on which I come 
back below. The EPR paper disputably failed to establish (*), even if one accepts its hypothesis and new concepts, because the arguments used a counterfactual. This is something that one should not put too hard  to Podolsky's charge because that concept was not mastered then.  In fact, Bell makes more serious errors of that sort  in his historical paper \cite{Bell} written 27 years later: see the critics to this effect in \cite{PenaCettoBrody1972}, a paper published in 1972 (see also the short account of  \cite{PenaCettoBrody1972} in \cite{Jammer1974}, p. 312, where \cite{PenaCettoBrody1972} is cited in its preprint form as \cite{J1972}).  Of course, the hypothesis of \cite{EPR} have also been severely contested, as I discuss at length in the present paper.

\smallskip
In Einstein's own words from 1936, taken from the reprinted form on p. 317 of \cite{EinsteinIdeasAndOpinions} (see also the much later text pp. 83-87 in \cite{Schilpp}):

\smallskip
[in the EPR entanglement situation]  ``Let us now determine the physical state of the partial system $A$ 
as completely as possible by measurements. Then quantum mechanics allows us to determine the
$\psi$ function of the partial system $B$ from the measurement made, and from the $\psi$ function of 
the total system [known since the initial $\psi$ function is assumed to be known, by
Schr{\"o}dinger equation]. This determination, however, gives a
result which depends upon \emph{which} of the quantities
(observables) of $A$ have been measured (for instance, coordinates
\emph{or} momenta). Since there can be only \emph{one} physical
state of $B$ after the interaction which cannot reasonably be
considered to depend on the peculiar measurement we perform on the
system $A$ separated from $B$ it may be concluded that the $\psi$
function in \emph{not}  unambiguously  coordinated to the physical
state".

\smallskip
\noindent 
From which (*) readily follows \textbf{if} one does not question the validity of the \emph{locality hypothesis} ({\bf 2}).

\smallskip
My own view is that locality, makes no physical sense at small enough scale, as will follow from the discussion in the last section. However, like Einstein, Omn{\`e}s \cite{Omnes}, and Griffiths \cite{Griffiths} for instance, I believe that it does hold true at the scale relevant for the EPR discussion, to the contrary of what is believed in Bell's camp (but I got the impression that the introduction of information theoretical concepts
- see, \emph{e.g.,} \cite {Fuchs0205039} - is slowly transforming some believers in non-locality into defenders of locality).  In effect, \emph{I believe that non-locality is a non-necessary departure from what is needed to make physics work, which furthermore introduces magical nuances that are contradictory to essential changes over Newtonian physics.}   

\subsection{3.3: A short nomenclature of instrumental realism}\label{sub:RealismNomenclature}
 
\begin{defin}[Realism and weak realism]\label{def:InstrumentalRealism}
In this paper,  \emph{realism} always means \emph{instrumental realism}, by this the following is meant:

\noindent
\emph{(Instrumental) realism holds true for some observable for some particle} if that observable gets a value and in particular pre-exists measurement on that particle, whenever that observable is measured on that particle, or the value of the observable converges to the measured value as one approaches from below the time of measurement.  

\noindent
\emph{(Instrumental) realism holds true in some setting} if it holds true for one at least of the particles that participate to that setting.  
\end{defin}
\begin{rem} [Realism without measurement] \label{rem;realismStand}
If realism holds true in some setting, I consider that it holds true as well when the measurement of the pre-existing measurable is not executed, but the rest of the setting remains the same. In some sense, this tells that future choices cannot change the present.
\end{rem}

Most physicists have come to adopt the point of view that realism does not work in microphysics. If one thinks about a particle involved in a two slits experiment, it seems clear that if one (suddenly) decides to measure the moment or position of that particle, that observable does not pre-exist measurement since otherwise one would get classical trajectories that would be incompatible with the interference fringes.  \emph{I will show in this paper (see Claim 1 in Subsection 4.1) that weak realism holds true.} 

\begin{rem} [On Einstein and realism] I want to emphasize that \emph{the text by Einstein from  p. 317 of \cite{EinsteinIdeasAndOpinions} (and reproduced above in subsection 3.2) stays clear from realism} and that \emph{factual reality} does not mean that observables have values before being observed, a point that will be discussed in Remark \ref{rem:RealismRevisited} and again, in the light of  Claim \ref{claim:C2}, in Remark \ref{rem:angels}.
\end{rem}

It has been solidly established, and often purposely ignored, that  the EPR paper \cite{EPR} has been written by Podolsky and was disliked by Einstein who most probably did not see the final version before it got printed (see in particular \cite{Jammer1974}, \cite{Jammer1985}, \cite{Jammer1989}, \cite{FineShaky}). Einstein's discussion is quite different, in particular where it comes to realism, from the point of view that one finds in the EPR paper \cite{EPR}.  So I propose a sort of scale of possible attitudes toward realism.  The attitudes that I list are not all expressed along the same point of view.  Anyway, the way one should think about realism in order to deal with microphysics is not that obvious, assuming even that there is in fact a single good angle, a simplification that is not clearly justified as far as I know. The list hereafter expresses my own views so that for instance ``ERP realism" means ``my own view about the ERP realism": I insist on making this quite clear because the subject 
has tremendously suffered from unfair attributions of believes and thoughts, and even from numerous miss-quotations.

\smallskip
\textbf{- (A) Strong realism:} \emph{All quantities pre-exist measurement}. Mermin's 3 particle GHZ setting allows one to show that this is inappropriate for microphysics (see Lemmas  \ref{lem:noY} and \ref{lem:noX}).

\smallskip
\textbf{- (B) EPR realism:} The EPR paper does not state explicitly a position on the existence of pre-existing value. We read there (as a definition of \emph{element of physical reality } \cite{EPR}:

\noindent
 \textit{``If, without in any way disturbing a system, we can predict with certainty (i.e., with probability equal to unity) the value of a physical quantity, then there exists an element of physical reality corresponding to this physical quantity."}

\noindent
\emph{I believe that, according to the EPR paper, 
quantities that can be predicted for sure pre-exist measurement}. At least this is something that can be read, I think, from Rosen's 1985 paper \cite{Rosen1985}, but since Einstein seems to not believe that, it is hard to know what was meant by Podolsky in \cite{EPR} (on Podolsky's aggressive attitude toward QM in 1935 when he wrote \cite{EPR} see, \emph{e.g.,} \cite{Jammer1974}).
\begin{defin}[Sensitivity to protocol and reallity]\label{def:ProtocolSensitivity}
Assume one has an element of reality with properties associated to measurements $A$ and $B$.
When one says that measurement $A$ has answer $a$ and measurement $B$ has answer $b$, it either mean means that:

- Measurement $A$ has answer $a$ if measurement $A$ is the one that is made and measurement $B$ has answer $b$ if measurement $B$ is the one that is made, in which case \emph{one has protocol sensitivity,}

\noindent
or that:

- Measurement $A$ would have answer $a$ and measurement $B$ would have answer $b$ whatever measurement is made or not, in which case \emph{one does not have protocol sensitivity.}
\end{defin}
The element of physical reality concept is utilized in the EPR paper using a counterfactual (see p. 779),  which tells us that the proof of non-completeness of QM in the EPR paper is not (fully) right (something hard to pin too hard on Podolsky as explained in Subsection 3.2) but this does not condemn the concept itself.  What is unclear is whether the ``certainty" in EPR's definition of elements of reality entails \emph{protocol sensitivity}, in the sense given by Definition \ref{def:ProtocolSensitivity}. 
This issue for historians is not discussed in the present paper (although I am not sure that historians have treated elements of reality fairly, so far); instead, I list possible points of view as follows: 

\begin{itemize}
\item \textbf{(B.1)} If certainty does not require protocol sensitivity (sensitivity upon the protocol being used to get certainty), we have what I call \emph{simple EPR realism}.  Mermin's 3 particle GHZ setting allows one to show (using Lemmas \ref{lem:noY} and \ref{lem:noX}) that simple EPR realism is inappropriate for microphysics, at least if one accepts locality. 
\item \textbf{(B.2)} If certainty requires protocol sensitivity, we have what I call \emph{protocol sensitive EPR realism}: the question remains open as far as I know as to whether this holds true in microphysics, or not.
\item \textbf{(B.3)} If  the \emph{simple EPR realism} is used only in non counterfactual ways, so that certain predictions are made on the basis of measurements on other particles that have been done or on the basis of a well defined unique measurement that could have been done (rather than on the basis of any experiment that could be done or of all experiments that could have been done), we have what I call \emph{counterfactual-sensitive EPR realism}: the question remains open as far as I know as to whether this holds true in microphysics, or not.
\end{itemize}
Elements of reality-based realism is usually attached to the name of Einstein (with an imprecision that has generated the list \textbf{(B.1)}, \textbf{(B.2)}, \textbf{(B.3)} just above) although Einstein did not use it when discussing the EPR setting.  Einstein even expressed that he did not care about the ``value of the observables" issue in a letter written in 1935 to Schr{\"o}dinger (see \cite{FineShaky}).  
\begin{rem} [Rosen's testimony on EPR and HV]\label{rem:RosenTestimony}
One is often induced to think EPR called for a very naive form of HV (the form used by Bell in \cite{Bell}), something which has been made highly disputable by a paper by Rosen \cite{Rosen1985}, written about EPR fifty years after the publication of the EPR paper.  Rosen also defends in \cite{Rosen1985} the soundness of the EPR paper if one accepts the hypotheses and views on QM expressed in \cite{EPR}, even if he then backs off from locality, an essential hypothesis of \cite{EPR}.  
\end{rem}

\textbf{- (C) Einstein's realism:} While Einstein's realism seem to have varied considerably, and some of his writings are hard to decrypt because of the allegorical tone, it is hard to pin more on him that the fact that \emph{real states are well defined anyhow}, which is what I will use as definition of  \emph{Einstein's realism} (or \emph{E-realism}) in the present paper. My opinion is that if any form of EPR realism had to be isolated as one Einstein would trust, it would be counterfactual sensitive EPR realism \textbf(B.3).

\smallskip
The above excerpt in Subsection 
3.2 , where Einstein gives his own rendition of the EPR issue, provides an example of how cautious Einstein was with matters such as realism and counterfactuals (even if some of the concept were not yet fully crystalized). What was clearly explained by Rosen, Fine, and others, is that reinterpreting Einstein's realism as meaning a belief in Hidden Variables of the form disqualified in a few lines by Bell has no foundation.  The only thing proven by such a reinterpretation is that Bell had a lot of chutzpah. 

\smallskip
\textbf{- (D) OQM-realism attitude:} Orthodox QM (see also Subsection 4.2) accepts that measured quantities \emph{post-exist} measurement. One can use here the following lines written by Nathan Rosen 50 years after he co-authored the EPR paper. 

\smallskip
\noindent
``[For Quantum Mechanics] The elements of reality are only those physical quantities for which the wave function predicts definite values. In the EPR example, if a measurement of A results in system II having a wave function $\psi _k(x_2)$ which is an eigenfunction of $P$, then $P$ is an element of reality. If measuring $B$ leads to II having a wave function $\phi _k(x_2)$ which is an eigenfunction of $Q$, then $Q$ is an element of reality.  Since there is no wave function of II for which both $P$ and $Q$ have definite values, they cannot both be elements of reality at the same time. According to this point of view, the description of physical reality by wave function is always complete because reality corresponds to what the wave functions describe."

\smallskip
\textbf{- (E) Weak realism:}  \emph{Weak realism} (or \emph{W-realism}) accepts, I recall, that measured quantities and quantities that have been predicted for sure do pre-exist measurement in some but not all cases.
 
\smallskip 
\textbf{- (F) E-W realism:} This is what one gets by combining weak realism with Einstein realism. This reflects my own position, but it is interesting to see what one can get from each of the E and the W.

\smallskip
\textbf{- (G) A-realism:} According to A-realsim, measured quantities do not pre-exist measurement. I hope that the present paper will show A-realism to be too extreme a position.

\begin{rem} [Jammer,  Einstein, and realism] 
While Jammer was most probably right to say that Einstein was the origin of this concept, the precise form of ``element of physical reality" in \cite{EPR} is different from what Einstein used in such context, a difference that is pointed out by Fine in \cite{FineShaky} and that can be checked on the few publicly accessible documents that one has from Einstein on such matters.
\end{rem}

\subsection{3.4: A few notes on \cite{EPR} and \cite{Bohm}}\label{sub:EPRReview}
The proof of (*) (or attempt of proof?)  given in \cite{EPR} is very nicely analyzed by Fine in \cite{FineShaky} although I may differ with him on assessing what (if anything) goes wrong 
with that proof. Briefly, Podolsky justifies two premises (to see clearly the structure that I report here following for instance \cite{FineShaky} or \cite{Barreau1985}, just analyze the paper starting from the end after reading it forward):

\noindent - the first one (see \cite{EPR} p. 778) is that:
\begin{itemize}
    \item (I-1): either (*) is true, or;
    \item (I-2): conjugate quantities cannot have simultaneous reality.
\end{itemize}

\noindent
- the second one  (see \cite{EPR} p. 780) is that:
\begin{itemize}
    \item (II-1): if (*) is false, then;
    \item (II-2): conjugate quantities can simultaneously
    have (arbitrarily) precise values.
\end{itemize}
From these premises, which Podolsky tries to establish, and QM, the statement (*) would follow easily. To prove these premises, a \emph{gedanken} experiment setting is proposed (which is of course the EPR setting described in Subsection 2.1), and Podolsky uses both locality and elements and reality, both of which have been charged later by some to be ``the problem".  This is because a lot of people got convinced (wrongly I believe as will be explained in Section 5) that Bell's enriched version of EPRB was leading to a contradiction as I recall below in Section 5 (while earlier attacks on \cite{EPR}, such as \cite{Bohr}  were at best not conclusive).  My view, is that the second premise of Podolsky is established
in \cite{EPR} using a counterfactual \emph{gedanken} experiment (another one, this time in the analysis of teh main one), which would make the paper wrong.  However, I am not sure of my attack because of the lack of clarity of the text which has kept other critics as perplexed as I am, even when the main issue that they have is different from mine.  

Besides the subtle (but I believe actual) use of counterfactual that I have detected in \cite{EPR} and that the reader is invited to discover as well directly from the source, I report here a clue of another sort that I consider as very convincing on the question of the counterfactual nature of the argument to prove (*) in \cite{EPR}.  This is the fact that in \cite{EPR}, Podolsky points out (in his own terms: the word counterfactual  is never mentioned in \cite{EPR} and the concept was even not yet isolated: imagine what Bohr could have answered in \cite{Bohr} otherwise, especially if Mermin and Peres for instance (and if I understand what they mean) are right on the fact that Bohr has attacked in \cite{Bohr} the counterfactual character of \cite{EPR}!) that if counterfactuals are not allowed, then his proof of non-completeness of QM breaks down.  More precisely, he writes:

\smallskip
\noindent ``Indeed, one would not arrive at our conclusion [\emph{i.e.,} (*)] if one insisted that two 
or more physical quantities can be regarded as simultaneous elements of reality \textit{only when 
they can be simultaneously measured or predicted.}  On this point of view, since either one or the other,
but not both simultaneously, of the quantities $P$ and $Q$ can be predicted, they are not simultaneously real. This makes the reality of $P$ and $Q$ depend upon the process of measurement
carried out on the the first system, which does not disturb the second system in any way. No 
reasonable definition of reality could be expected to permit this."

\smallskip
\noindent Hence Podosky also says (again, in his own terms) that elements of reality that would 
need to adapt to counterfactual analysis cannot make sense: he thus missed a deep difficulty 
that comes with talking about Quantum Mechanical phenomena, and created for himself a trap in which Einstein did not fall in his own further presentations.

\smallskip
As for the attacks by Bell, I try to dismiss them in Chapter 5, also on the basis of the use by Bell of counterfactuals, something to which Einstein escaped in his treatment of the EPR problem has reported above in Subsection 3.2. I hope to come back on these issues elsewhere but I do not want to dive here into too much of historical digressions. 

\medskip
Bohm tried in his version \cite{Bohm} of EPR, to call the attention on the fact that, when using the singlet state and measuring spins on axes $x$ and $y$ that are mutually orthogonal and orthogonal to the axis of propagation, one get something that resembles action at a distance since for any choice of $(x,y)$, and assuming that the first measure is performed on the electron and along $x$, one gets for the positron: 

- Minus the electron reading if one performs an $x$-reading,

-  Equidistribution of the outputs + and - readings if one performs a $y$-reading.

It is this discussion (or EPRB as reiterated in \cite{BohmAharonov1957}  in the framework of photon polarization) that got enriched in Bell's work to be reviewed briefly in Chapter 5.  I hope to come back on these issues elsewhere as well but like in the case of the original EPR above, I  want to avoid here further historical digressions. Besides, I will provide in \cite{Tresser2005F} a simpler counter-argument to the paradoxical aspect of EPRB that what I achieve in the present paper as a byproduct of Claims \ref{claim:C0} to \ref{claim:C1'}. 
%
%
\section{4: Conservation laws and Schr{\"o}dinger's  Unorthodoxy Theorem}\label{sec:conservation}
\subsection{4.1: Conservation laws and instrumental realism: the first main EPR claim}
The point of view that conservation laws cannot be invoked in the EPR setting has become quasi official, despite the counterfactual character of Bell inequalities, or rather because that character has been too often overlooked. Refusing the role of conservation laws in the EPR setting goes in the same direction as lending to Einstein naive points of view that are easily shown to be wrong.  The following Claim that helps me defend the opposite position.  It turns out that this claim is essential for most of what is presented here. 

\begin{claim}\label{claim:C0}
In an EPR-type situation (with a symmetric pair of particles or not), assume that one performs on one of the particles the measurement of an observable corresponding to a conserved quantity once the particles are separated.  Then the outcome of the measurement and the conservation law corresponding to the conserved quantity allow the knowledge of the value on both particles of that observable. In particular,  instrumental realism holds true in this situation, from which the rest of the claim follows readily.
\end{claim}

\begin{rem} [Quasi invariance] \label{rem:larry}
As pointed out to me by Larry Horwitz, all conservation laws need not remain absolute in QM.   For instance, the constraint of the total spin remaining constant along an axis (\emph{e.g.,} staying equal to zero along any axis in the case of the creation of a pair) may have to be taken with a grain of salt because of a phase problem when the particles separate, at least in the non-relativistic case that we consider here. But that would not affect the discussion as quasi-conservation laws can be used as well as long as they are effective laws and the quasi-conservation is absolute and not merely statistical or of probability one. We also implicitly assume that curvature effects are negligible.
\end{rem}

\begin{rem} [Two-States and Many Worlds]  \label{rem:Lev}
Important schools of interpretation of QM, including Yakir Aharonov, Lev Vaidman and some of their collaborators, have developed a view of the world that includes the Many World Interpretation of QM, which indeed allows to reduce to nothing many phenomena that appear as paradoxes in other interpretations of QM. The arsenal of tools being used include the use of a two-state vectors formalism, and the points of views taken include: 

- $\alpha$) considering that QM is nonlocal

\noindent
and:

- $\beta$) considering that if two conjugate observables are measured on a particle ${\bf m}$ at times $t_1$ and $t_2$ with $t_2>t_1$, with respective values $\pi _1$ and $\kappa _2$, then if no interaction involving ${\bf m}$ happens in the open interval $(t_1,t_2)$, in that interval the second observable has the value $\kappa _2$ (then the other observable has value $\pi _1$ but no one would have an issue with that).

\noindent
The point of view $\beta$ is what I would call (systematic) instrumental realism, according to the nomenclature proposed in subsection 3.3: this clashes with Remark \ref{rem;realismStand} except if one accepts strong realism as defined in subsection 3.3.

There is an important body of work behind the few lines of the present remark: l refer to the review paper \cite{AhaVaidmanReview}, and just mention \cite{AhaVaidAlb} as a relatively early paper where the instrumental realism in the form of $\beta$ is explicitly stated.  
\end{rem}

\noindent \textbf{Arguments for Claim \ref{claim:C0}.} 
The use of separation is trivial, with details left to the reader. By conservation of the quantity at hand (which is commented upon in Remark \ref{rem:larry}), if the initial value of that quantity is $v_0$ and the measurement on particle 1 gives $v_1$, then the value of the same observable on particle 2 is $v_2=v_0-v_1$ if the measure on the second particle is also performed (or at least approximately so: see Remark \ref{rem:larry}). 

\smallskip
\textbf{I). Conservation law argument.} At any time before measurement on the second particle, the value of the same observable on particle 2 is $v_2=v_0-v_1$ whether or not the measurement will eventually be made or not, since the total value of the observable is conserved (or at least approximately so: see Remark \ref{rem:larry}).

\smallskip
\textbf{II). WPR argument.}  One is now invited to consider WPR as a mathematical trick.  Anyway, WPR or not, the packet is reduced after the first measurement (on the first particle) so that there is a true value to be ``read" on the second particle in case the second measurement is performed.

\smallskip
\textbf{III). Secure value prediction argument.} I notice that, if $v_0$ is known (which is often the case with $v_0=0$ indeed), the result $v_2=v_0-v_1$ appears as a sure measurement prediction either by using the conservation law, or by using WPR which makes $v_2$ appear as an eigenvalue for the operator corresponding to a measurement of the observable on the second particle (as described in fact in \cite{EPR}). This secure prediction of the value can be considered as extension of the definition of the value: experimental verifications could be made if the position and momentum could be observed since measuring $\vec{p}_1$ and inducing the other moment $\vec{p}_2$, after measuring the position $\vec{q}_2$ and taking the pair $(\vec{p}_2, \vec{q}_2)$ as an arbitrarily good approximation close enough to where $\vec{q}_2$ is measured (see also Claim \ref{claim:C1} and the arguments for it), one could compute trajectories for the two particles.

\begin{rem} [About \cite{Pitowsky} and GHZ]\label{rem:Pitowsky}
In \cite{Pitowsky} one can find what seems to be a direct contradiction to the Secure value prediction argument. However, the authors consider a case with no measurement made in the context of GHZ. When one of the observers considers that an outcome is sure, it is in fact sure if some corresponding measurements and not other ones are performed by the two other observers. With no counterfactual, there is no problem in the GHZ situation. If an observer knows the result of his or her observation, and the protocols to be used by the two others, the product of the values to be read by the two others is known for sure. Similarly, if two observers have made their observations and can communicate, then if they know the protocol to be used by the last observer, they can predict surely the outcome of the third measurement. The two last statements can be reformulated to cover the case when one observer, or two observers who can communicate, consider what can be deduced from the different possible outcomes of the measurements that each of them could perform. 
\end{rem}

Notice that as long as the prediction of values is made on the basis of \emph{other measurements that have been done} rather than on the basis of \emph{other measurements that could be done}, the Mermin GHZ setting does not any more provide a counter-example to the ``Secure value prediction argument". In fact, neither the protocol sensitive EPR realism \textbf{(B.2)} nor the counterfactual-sensitive EPR realism  \textbf{(B.3)} are faulted by the Mermin GHZ setting as a careful examination confirms. 

\medskip
The Conservation law and the Secure value prediction arguments could be disputed by some as being manifestations of realism as usual. This would be wrong, however I next provide an alternate argument based on the interpretation of some ``one pair at a time" experiments.

\smallskip
\textbf{IV). Experiments interpretation argument.} In order to argue on the basis of experiments, I will need to involve the more quantitative Claims 2 and 3 that are shown later to be easy consequences of Claim 1, and even possibly some further tests discussed in Remark \ref{rem:academic+} in the case of momentum and position.  This is perfectly fine as the goal is to get a coherent theory for the physical world, and not at all to build an axiomatic theory (the hope of catching the description of the world into an axiomatic theory is still amazingly strong, but is probably an anthropocentrism awaiting its Copernician Revolution; I rather think that it is mostly the fossilized pieces of science that admit a full mathematical description).

\smallskip 
In experiments like \cite{eraser} and \cite{eraser2}, one uses one element of an EPR pair to perform a double slit experiment or another comparable experiment.  Performing a position (or vector momentum) measurement on the other particle (say the second particle) allows us to know enough of the trajectory of the first particle to determine which slit would be used.  This explains that \emph{as an experimental fact there are no interferences in one particle at a time double-slit experiments performed with one of the members of an EPR pair.} The same suppression of interferences on the first particle arises even if no measurement is performed on the second particle. Hence, as a sign of locality by itself, \emph{performing or not the measurement on the second particle of an EPR pair does not change anything to the fact that the path of the first particle is predictable and does not produce interference fringes}.  I interpret this suppression of interference fringes as the fact that the EPR type entanglement selects the particle nature of the elements of a pair, just as measurements (which formally are like entanglements), often select one or the other of the particle or wave characters.  The measurement-like character of entanglements is a well known aspect already mentioned as far back as \cite{Schrodinger1935a}.  Once trajectories make sense, a small act of faith seems to be needed to accept that the momentum preexists measurement, as well as position (where position values approach measured values along the trajectories):  

- Either we will need to build on the consequences of Claim \ref{claim:C0} as expressed in Claim \ref{claim:C1} to get to means, in the form of Remark \ref{rem:academic+}, to experimentally support the fact that indeed realism holds true for the momentum and for the position of the two particles in the EPR context.  More precisely Claims  \ref{claim:C0} to  \ref{claim:C1} together with the experiments that we have so far and with the experimental tests that can be built following Remark \ref{rem:academic+} are expected to form a coherent combination of theory and data.   

- Or we admit that the only impediment to classical coordinate is the lack of trajectories, so that (\emph{de facto}) experimental evidence for trajectories is (\emph{de jure}) experimental evidence for position and momentum.

\smallskip
A gap needs to be bridged when extending the validity of instrumental realism to observables that are not involved in the real trajectory-like behavior in double slit and similar experiments.
However, in the case of the polarization of photon (formally similar to the spin of spin-$\frac{1}{2}$ particles), one can invoke the experiments that were performed \emph{a priori} to verify that QM does not satisfy the Bell inequalities.  These experiments  (see, \emph{e.g.,} \cite{AspectEtAl1982},  \cite{Innsbruck1998}) do certainly not achieve what is claimed that they do because of the counterfactual nature of Bell type inequalities.  However they do offer further verifications of QM and in particular of the Malus Law up to the sign (so much that we will not need the counterpart of Remark \ref{rem:academic+} in the spin case). 

\begin{rem} [Malus Law] \label{rem;Malus}
I briefly review Malus Law for spin-$\frac{1}{2}$ particles measured along two successive Stern-Gerlach magnets respectively along  $\vec{a}$ and $\vec{b}$,  that both generate +1 for a positive spin, -1 for a negative spin along their respective vectors.  Malus Law reads 
\begin{equation}\label{MalusP}
\mathcal{Q}(\vec{a},\vec{b})=\vec{a}\cdot \vec{b}\,,
\end{equation}
when expressed in terms of \emph{the average of the product of the successive spins along $\vec{a}$ and $\vec{b}$}. One also often uses 
 $\mathcal{S}(\vec{a},\vec{b})$, \emph{ the probability of having the same reading}, which is
related to the other  quantity by $\mathcal{Q}(\vec{a},\vec{b})=2\mathcal{S}(\vec{a},\vec{b})-1$.  With $\theta$ standing for the angle between $\vec{a}$ and $\vec{b}$, some readers are probably more familiar with the formula 
\begin{equation}\label{MalusS}
\mathcal{S}(\vec{a},\vec{b})= \cos ^2(\frac{\theta}{2})\,,
\end{equation}
and use Equation \ref {MalusS} as the expression of the Malus Law  rather than  Equation \ref {MalusP} which can be rewritten as 
\[\mathcal{Q}(\vec{a},\vec{b})=2 \cos ^2(\frac{\theta}{2})-1= \cos (\theta)\,.
\]  
\end{rem}

Coming back to the experiments (supposedly) related to Bell's inequality (a matter that will be discussed further in Section 5), one uses there Stern-Gerlash magnets respectively along  $\vec{a}$ and $\vec{b}$, respectively for the two particles of an EPR pair (in fact, one uses polarizers since the experiments are performed with EPR photons, but the algebra is identical). 
The sign for the average of products $\mathcal{P}(\vec{a},\vec{b})$ of spins from a pair (or of polarizations of photons from a pair)
gets reversed with respect to the Malus Law, so that QM predicts (see Section 5): 
\begin{equation}\label{MinusMalusP}
\mathcal{P}(\vec{a},\vec{b})=-\vec{a}\cdot \vec{b}
\end{equation}
Thus, by confirming Equation \ref{MinusMalusP}, the experiments aimed at supporting verifying that Nature follows QM rather than very naive Bell-type inequalities fail their goals (doubly as neither QM nor nature accepts the counterfactual needed in Bell-type inequalities) but provide a strong confirmation of instrumental realism.  More precisely this confirmation:

- Is obtained directly for the polarization of photons, in view of the analysis made in the arguments for Claim \ref{claim:C1'} below, so that Claims  \ref{claim:C0} to  \ref{claim:C1'} together with the experiments done so far form a coherent combination of theory and data.  

- Requires some good faith to be extended to the spin of spin-$\frac{1}{2}$ particles, given that the algebras for the spin of spin-$\frac{1}{2}$ particles and for the polarization of photons are alike while no experimental data are available in the case of the spin of spin-$\frac{1}{2}$ particles.

\smallskip
Summarizing now what we have learned about several observables: 

- Experiments of the \cite{eraser} or \cite{eraser2} type (contribute to) provide evidence of the fact that instrumental realism works on the second particle, hence on the first particle as well by symmetry of labeling for momentum and position.  

- Experiments aimed at checking the status of Nature as compared with what is told by the Bell inequality as in \cite{AspectEtAl1982},  \cite{Innsbruck1998} cover the case of the polarization for photons, from which I feel comfortable for the case of the spin of spin-$\frac{1}{2}$ particles as well. 

\noindent
One might consider that there is then no good reason to not believe that instrumental realism also works for other observables.  Anyway, we have more than enough to allow us later to prove Claim \ref{claim:C1}.

\smallskip
\noindent
The instrumental realism is the hard part of Claim \ref{claim:C0}: the rest follows easily since we know what happens when measurements are made on the two particles.  This finishes the arguments to establish Claim \ref{claim:C0}.

\begin{rem} [Entanglement and measurement] \label{rem:EntangMeasure}
The rationale for instrumental realism to work in all cases is that entangled particles in EPR pairs sort of ``measure each other". It seems reasonable to consider the wave-particle dualism as the main cause of non-realism in general, and EPR entanglement as a dualism breaker. The formal identity of EPR entanglement and measurement goes as far back at least as 1935 \cite{Schrodinger1935a}.
\end{rem}

\medskip
It so happens that the term \emph{eraser} is used in the context of the experiments that I consider here, but to describe another effect than the suppression of interferences that I have invoked in my argument: the interference fringes are restored and then \emph{erased} (again, so to speak) by further manipulations.  Eraser experiments show that entanglements are really two particle states, and that discarding this fact prevents one from understanding some of the phenomenology.  This does not contradict the fact that, whenever the two particles are far apart, some observables can make sense for each of them, which is all that is needed in order to justify the point of view that I defend. In fact, all my arguments hinge in some way on the very special character of entangled states of the EPR type, as did Einstein's argument and the argument presented by Podolsky in \cite{EPR}. 

\begin{rem}[Is weak realism new?]\label{rem:CheapRealism}
The very concept of weak realism (instrumental realism special to some settings, such as EPR entanglements I recall) might be new in explicit form.  However, it seems obvious that the EPR authors and Schr{\"o}dinger at least felt that something like this was true since 1935, 69 years before the present work began.  Measurement and new developments in Quantum Computing make entanglements a rather pervasive phenomenon, as Schr{\"o}dinger already noticed in \cite{Schrodinger1935a}. This does not mean that instrumental realism is prevalent since the GHZ setting clearly shows that some entanglements do not generate pre-existence to experiments of observable values. 
\end{rem}

\subsection{4.2: Measuring on both EPR particles: the second main EPR claim}
\begin{claim}\label{claim:C1}
In an EPR-type situation, one can access with arbitrary precision the values of both members of a pair of conjugate observables on at least one of the elements of the pair of particles if instrumental realism holds true for both observables.  Assuming that the measurements on the two particles are space-like separated, this access to information takes place after the second observable is measured on the second particle.  The information so obtained on values of a pair of conjugate observables 
is about the second particle.  This information is valid before that second measurement on the second particle but after the first measurement has been performed on the first particle.
\end{claim}
This claim is not new in essence but this explicit form is both weaker and broader than the original result 
by Schr{\"o}dinger in \cite{Schrodinger1935a} (see \cite{Jammer1974}, \cite{Fine2}). What seems also to be new is the non-trivial time shift about when double knowledge is valid and when it is made available.  See also the series of papers \cite{Schrodinger1935b} where Schr{\"o}dinger  furthermore introduces the celebrated \emph{Paradox of the cat} which grew out of correspondence with Einstein (see \cite{FineShaky} for a critical review of this correspondence). 

\noindent
I pause here to formulate a definition that will help map the result of Schr{\"o}dinger in \cite{Schrodinger1935a} and \cite{Schrodinger1935b} to some of the preoccupations of Einstein.

\begin{defin}

- (i) \emph{Orthodox Quantum Mechanics}, as I define it, postulates simultaneous inaccessibility to conjugate observable, even by means not previously thought about, so that the Hilbert spaces and Hermitian operators acting on them are perfectly matched by the theory that describes the physical world.

- (ii) Orthodox Quantum Mechanics also rejects the validity of (*).  This rejection is exactly Einstein's 
own characterization of Orthodox Quantum Mechanics: on p. 681 of \cite{Schilpp}, he writes: 

``\emph{``orthodox" refers to the thesis that the $\psi$-function characterizes the individual system} exhaustively."

I have extended Einstein's definition because of what I feel is more basic for modern QM, while Bell's arguments for non-locality revolve around the negation of (*). In what follows, Orthodox means either (i),  (ii), or both, depending on the context.  For instance I will establish the violation (i) below in this subsection as a way to correct Podolsky's argument in \cite{EPR}: the violation of (ii) then follows. See also below, Remark \ref{rem: Orthodox}.
\end{defin}

Schr{\"o}dinger reaches in \cite{Schrodinger1935a} the conclusion that there can be too many observablesÕ values \emph{per} particle in the EPR configuration by measuring say, the position $\vec{q}_1$ on particle 1 and the momentum $\vec{p}_2$ on particle 2, and inferring from that the position $\vec{q}_2$ of particle 2 and the momentum $\vec{p}_1$ of particle 1.  
From there he gets into issues of comparing values of functions of observable with functions of values of observables, which arguably lead to potential difficulties (see \cite{Schrodinger1935a} and \cite{Fine2}). 
The key remark behind that, indeed one of the well known EPR facts, is that 
$\vec{q}_1-\vec{q}_2$ and $\vec{p}_1+\vec{p}_2$ commute. I come back to the algebra after the following remark.
\begin{rem} [Asking for less] \label{rem:Ask4less}
Also well known, but not taken into account by Schr{\"o}dinger, is the remark that the origin, the place where the particles are starting from, is not (precisely known). Because of that the Uncertainty Relation would take its toll right there if need was of determining one of $\vec{q}_2$ and $\vec{q}_1$  from  measuring the other one. 
In fact, instead of $\vec{q}_2+\vec{q}_1=\vec{0}$, we have $\vec{q}_2+\vec{q}_1=\vec{U}$, 
where $\vec{U}$ stands for some unknown vector.
In what follows, I just try to get the values of both 
observables on one of the particles, except if a second conservation law allows for more, as it is the case with spins in the EPRB configuration. Thus there appears to be a flaw in Schr{\"o}dinger's approach,  but that flaw gets corrected in the argument that I provide below to establish Claim \ref{claim:C1}, mostly by asking for less and changing:

- The time when the simultaneous values of two conjugate observable on one particle makes sense, 

\noindent
and

- The time when such knowledge becomes accessible. 
\end{rem}
The EPRB configuration is the framework of the short paper \cite{Peres1990} where Peres focuses on the algebraic play between values of functions of observable and functions of values of observables.  Without any mention to \cite{Schrodinger1935a} nor \cite{Schrodinger1935b}, Peres is trying in  \cite{Peres1990} to 
get even stronger contradictions than in \cite{EPR} out of the use of the concept of element of reality  (see also p. 151 in his book \cite{Peres1993}).  The goal of\cite{Peres1990} is (as I understand it) to get stronger  arguments than before against what, like others, he presents as \emph{``the local realism of Einstein"} (see Subsection 3.3 above and references therein for alternates views on that).  
Peres takes seriously enough the algebraic contradictions that he can reach to claim derailing, at least implicitly, the EPR attempt at questioning QM by raising the issue (*).  However, he uses a lot of counterfactuals in \cite{Peres1990} (and in \cite{Peres1993}) to reach a contradiction (see also the analysis of Peres' argument in \cite{Fine2}).
Thus the contradiction reached by Peres, under the assumption that one gets values of two conjugate spin observables on the two particles of the EPRB pair, is moot and does not compromise the validity of arguments to establish Claim \ref{claim:C1}, and even less the validity of Claim \ref{claim:C1} itself. Notice that formulas for products of spins, one of which is obtained by instrumental realism, have 
no reason to be related to products of spins actually measured consecutively on the same particle, which are the products that one finds in most textbooks on QM.  In brief, the anti-realist conclusions that are reached in Peres' paper are comparable to, and can be argued against as, the conclusion of Mermin's version of the GHZ entanglement (a version of which appears on pp. 152-153 of \cite{Peres1993}).  

\begin{rem} [Beyond academic value] \label{rem:academic+} 
Getting the values of two projections of a spin of a particle before that particle gets measured for one of the projection may seem to be a candidate for one more ``so what?" among the QM-related claims that have no experimentally verifiable content. However, in the $(\vec{p},\vec{q})$ case, in the academic case when $\vec{q_1}+\vec{q_2}$ is known, one would be able 
to compute actual trajectories by using (say) $\vec{p_2}=-\vec{p_1}$ and the measured value of $\vec{q_2}$ as an arbitrary good approximation of $\vec{q_2}$ just before that measurement. In fact, one can also do that in the actual, non-academic case when the value of the conserved quantity $\vec{q_1}+\vec{q_2}=\vec{U}$ remains unknown, but then, to the contrary of the academic case, one cannot then describe as well the trajectory of the first particle.   
\end{rem}

\noindent \textbf{Arguments for Claim \ref{claim:C1}.}

- $\alpha )$ Notice first that Claim \ref{claim:C1} is plain in the case of the (energy, time)
pair or $(E,\, t)$. This case seems to not need entanglements, and this would 
be the case if one could find a process that generates particles with known fixed energies.  This is only one aspect of the fact that the status of the $(E,\, t)$ pair.  Time measurement does not correspond to a Hermitian operator and the status of this pair with respect to the Uncertainty Principle is special as commented upon already in  \cite{LL1958}.  In fact (see \cite{LanPe1931}, \cite{LL1958}, \cite{AhBo1961}, \cite{Fock1962}, \cite{AhBo1964}, \cite{Peres1993} (pp. 413-415),  \cite{AhMaPo} and references therein):
 
- If the Hamiltonian $H$ is not known, then a measurement necessarily takes a minimum time $t$ which 
obeys an uncertainty relation of the form $\Delta t\cdot \Delta E \geq A$, where $A$ is a number that depends on the Hamiltonian $H$ and where $\Delta E$is the precision of the energy measurement.
 
 - If the Hamiltonian $H$ is known then it is claimed in \cite{AhMaPo}  (see also \cite{AhBo1961}, \cite{AhBo1964}) that we can measure the energy as precisely as we want in a time as short as we want. However (as noticed in \cite{Fock1962}), in order to refute the  $(E,\, t)$ uncertainty relation Aharonov and Bohm introduce into the Hamiltonian in \cite{AhBo1961}, an interaction term involving a discontinuous function of time. This procedure implies that instantaneous energy changes of a predictable amount and at a given instant can be observed, which is itself in violation of the uncertainty relation. Aharonov and Bohm's 1961 argument thus contains a \emph{petitio principii}, which is corrected in \cite{AhBo1964} by letting the discontinuity happen in space. Peres defends in \cite{Peres1993} (pp. 413-415) that QM is a construct like Euclidean geometry (for instance) and should be judged internally as such, so that the possibility to realize a given setting as in \cite{AhBo1964} is of practical and not of fundamental importance. This view clearly clashes with the one that I defend here, that QM is a physics theory that has suffered from premature (and quasi-religious) transformation into an axiomatic theory.  Peres also mention (on p. 415 of his book)  the other Uncertainty Relations exclusively in their dispersion meaning, but to my opinion, this is providing a quite interesting but very partial view on a multi-faceted phenomenon.  Hence, since such matters will not have too much consequences on points $\beta )$,  $\gamma )$, and $\delta )$, and given the motivations that I have which need only one instance where the axiomatic formalism of QM would be chattered, I stop short here on the discussion of point $\alpha )$.  I leave to the reader to decide if the case can be made with the $(E,\, t)$ pair and turn my attention to other pairs of conjugate observables.
  
\smallskip
- $\beta )$ For the (momentum, position) pair or $(\vec{p},\vec{q})$, we have seen in Remark \ref{rem:Ask4less} that in general, the vector  $\vec{q_1}+ \vec{q_2}$ is conserved to an unknown value, say $\vec{U}$ not necessarily equal to $\vec{0}$ .  This does not prevent however the arguments for Claim \ref{claim:C0} to work as well as when $\vec{U}=\vec{0}$.  Thus instrumental realism holds true for the positions $\vec{q_1}$ and $\vec{q_2}$, and we can take the measured value of $\vec{q_2}$ as an arbitrarily good approximation of what is $\vec{q_2}$ just before the measurement takes place at time $t_2$ on particle 2.  Furthermore, using Claim \ref{claim:C0}, I also get $\vec{p_2}$ as $-\vec{p_1}$, where $\vec{p_1}$ has been measured on particle 1 at time $t_1$, before (in any frame, because of the space-like separation hypothesis) measuring on particle 2. So after $t_1$ and just before $t_2$, I get $p_2$ and almost $q_2$ (with arbitrarily good approximation) on particle 2.  This is less than what Schr{\"o}dinger got in \cite{Schrodinger1935a}, which might be an advantage, as pointed out  in Remark \ref{rem:Ask4less}.

\smallskip
- $\gamma )$ Still for $(\vec{p},\vec{q})$  the claim can also be applied to the (academic) case when $\vec{q_1}+ \vec{q_2}$ is conserved to a known value.  In that case, one falls in the general case of two conserved quantities with known initial values (the associated difficulties, like those that Schr{\"o}dinger got in \cite{Schrodinger1935a}, have already been commented upon: see for instance Remark \ref{rem:Ask4less}).

\smallskip
- $\delta )$ We now get to the general case  of two conserved quantities with known initial values,  that form a pair of conjugated observables, such as for instance the spin projections along two orthogonal axes, say $x$ and $z$.  On an horizontal axis $y=(-\infty, "near"\,\, O]\cup["near"\,\, O, \infty)$ that represents approximately the two trajectories joined at or "near" $O$  ($O$ is known approximately, with some reasonable bounds: for instance one knows where is the macroscopic apparatus producing the pairs), let us mark two further points (possibly up to some small mistake):

- $L$ to the left of $O$ (far enough),

- $R$ to the right of $O$ (far enough).

\noindent  For definiteness, the electron ${\bf e}$ is going to the left and the positron ${\bf p}$ is going 
to the right.  Measurements will be made on ${\bf e}$ at (or near) $L$, and on ${\bf p}$ at (or near) 
$R$. Notice that any conserved quantity remains unchanged on $("near"\,\, L,\, "near"\,\, O)$ and on $("near"\,\, O,\,"near"\,\, R$, with possible sign reversal at (or near) crossing $O$ (for such matters related to conservation laws, see however Remark \ref{rem:larry}). 
 
However, inferring this fact (then in view of time invariance) may appear to be a realistic view that should be subject to question if not rejected right away if the conservation on $("near"\,\, L,\,"near"\,\, O)$ (respectively $("near"\,\, O,\,"near"\,\, R)$) is inferred to claim that the measured value at  or near $L$ (respectively at or near  $R$) is meaningful as well on $("near"\,\, L,\,"near"\,\, O)$ (respectively $("near"\,\, O,\,"near"\,\, R)$).  However, for conserved quantities, one can invoke the corresponding conservation law  and Claim \ref{claim:C0} to infer the conservation of the value of the observable - up to sign - on $("near"\,\, L,\,"near"\,\, O)$ (respectively $("near"\,\, O,\,"near"\,\, R)$) from the measured value at or near $R$ (respectively at or near $L$).  Then one can invoke again Claim \ref{claim:C0} and the very same corresponding conservation law to infer, up to sign, the value of the observable on the other of the $("near"\,\, L,\,"near"\,\, O)$ or $("near"\,\, O,\,"near"\,\, R)$ segments than the one that has been used so far.  

This being said, I will always rely on this discussion when I allow myself to skip writing it and say 
globally that \emph{any conserved quantity remains unchanged on $("near"\,\, L,\,"near"\,\,O)$ and on $("near"\,\, O,\,"near"\,\, R)$, with 
possible sign reversal at (or near) crossing $O$}.  

The fact that ${\bf e}$ looks like \emph{``minus the particle ${\bf p}$ {\bf after} the measurement has been made on $("near"\,\, L,\,"near"\,\,O)$"} rather than  (in fact as well as) \emph{``minus the particle ${\bf p}$ {\bf before} the measurement has been made on $("near"\,\, L,\,"near"\,\,O)$"} is linked to the special entanglement structure which makes some special conclusions of the realist point of view work fine (see Claim \ref{claim:C0}).

\smallskip
\noindent 
I present only the case of the spin for clarity, with no serious loss of generality.  Thus assuming that 
the direction $x$ is used for measuring ${\bf e}$ at (or near) $L$ yielding $Q_1=s_x({\bf e})$, and 
that the direction $z$ is used for measuring ${\bf p}$ at (or near) $R$ yielding $Q_2=s_z({\bf p})$, 
one notices that ( with multiple use of Claim \ref{claim:C0}):

\begin{itemize}
\item $Q_1$ is conserved on $["near"\,\, L,\,"near"\,\, O]$, so that $\tilde{Q}_1=s_x(p)=-Q_1$ is conserved on $["near"\,\, O,\, "near"\,\,R]$. 
\item $Q_2$ is conserved on $["near"\,\, O,\,"near"\,\, R]$, so that $\tilde{Q}_2=s_z(e)=-Q_2$ is conserved on $["near"\,\, L,\,"near"\,\, O]$. 
\end{itemize}

Thus in the case of conjugate variables, measuring on both ${\bf e}$ and ${\bf p}$ allows simultaneous knowledge of this pair of conjugate observables on the open intervals before measurement and far enough from $O$ on both sides, which ends the arguments for Claim \ref{claim:C1}.

\subsection{4.3: Some more comments on Claim \ref{claim:C1}}
As pointed out to me by Marco Martens when I told him about 
Claim \ref{claim:C1}, the possibility to measure, one on each side, observables that are
conjugate becomes plain when one realizes how obvious it is that these measurements commute. This commutation property is equivalent to Claim \ref{claim:C1} once given Claim \ref{claim:C0}: indeed this yields another (shorter) proof of Claim \ref{claim:C1}, using again Claim \ref{claim:C0}. The main advantages of the proof that I have chosen are:

- the fact that it distinguishes the case of two conserved quantities, so that the proof proves more than what is stated.

- the fact that this proof adapts well for the needs of Claim \ref{claim:C1'}, which was the original motivation for Claim \ref{claim:C1}. 

\noindent
However, as far as the statement chosen for Claim \ref{claim:C1} is the sole concern, the commutation property-rule-based proof is perhaps the best one, and it reduces the realism issue to simply using Claim \ref{claim:C0} and choosing a good time to get the values of the two observables.
\begin{rem}[The dent in Orthodox  QM]\label{rem: Orthodox}
Remark \ref{rem:CheapRealism} 
has begun discussing the dent in Orthodox 
QM made by Claims \ref{claim:C0} and \ref{claim:C1}. By Claim \ref{claim:C1}
joint inaccessibility to conjugate variable appears as \emph{generic} (instead of \emph{general}).  However, joint inaccessibility is put, I think, on  a less magical footing by Claim \ref{claim:C1} than by axiomatic views, either Orthodox or Consistent (see \cite{Griffiths}, \cite{Omnes}).  Thus the reason for joint inaccessibility  now appearsto be the \emph{Uncertainty Principle} as formulated for 
measurements effectively made, (\emph{e.g.,} on simple isolated systems).
\end{rem}
%
%
\section{5: Two aspects of Bell's 1964 paper} \label{sec:Bell}
\subsection{5.1: The two aspects of Bell's 1964 paper}
Except for the Introduction, which has too often escaped comments (see however \cite{Jammer1974}) and that I will discuss elsewhere (but see also Remark \ref{rem:BadBoy}), I will mainly break down the content of Bell's historical 1964 paper \cite{Bell} in two parts:

\smallskip
\noindent
\textbf{Part 1 where Bell:}

\noindent - Introduces the idea of using an arbitrary 
angle $\theta$ between two Stern-Gerlach vectors $\vec{a}$  and $\vec{b}$ that respectively define $s_{\bf e}=+1$ for the electron and $s_{\bf p}=+1$ for the positron, 

and then

\noindent - Reports that Quantum Mechanics predicts
$-\vec{a}\cdot \vec{b}$ for the average $\mathcal{P}(\vec{a},\vec{b})$ of $s_{\bf e}(i)\cdot s_{\bf p}(i)$ when the pair of Stern-Gerlach vectors $(\vec{a},\vec{b})$ along which measurements are made is  is chosen, with $i$ standing for the $i^{\rm th}$ pair $({\bf e},\, {\bf p})$ in the singlet state that is being tested (see Remark \ref{rem;Malus} and the few line thereafter).

\smallskip
\noindent
\textbf{Part 2 where:}

\noindent  Bell first makes us believe that \emph{for a local and predictive Hidden Variables theory with the same statistics as Quantum Mechanics}, 
the average $P(\vec{a},\vec{b})$ of the product of the spin of {\bf e} along $\vec{a}$ by the spin of {\bf p} along $\vec{b}$ would be the quantity:
\[
P(\vec{a},\vec{b})=\int [s_e(\vec{a},\lambda)\cdot s_p(\vec{b}, \lambda)]\, \rho (\lambda)\,d\lambda\,,
\]
with $\lambda$ standing for the (possibly multidimensional) hidden parameter assumed to have an integrable density and $\rho(\lambda)$ the normalized density.  From that formula and the explicit and implicit hypotheses leading to it,  \emph{Bell easily deduces:}
\[
1+{P}(\vec{b},\vec{c})\geq|{P}(\vec{a},\vec{b})
-{P}(\vec{a},\vec{c})|\qquad (**)\,.
\]
In  fact, in view of  $s_e(\vec{b},\lambda)= -s_p(\vec{b}, \lambda)$ and noticing that 
$s_e(\vec{b},\lambda\in  \{-1,\,+1\}$, it is enough to remark that:
\begin{widetext}
$$
P(\vec{a},\vec{b})-P(\vec{a},\vec{c})
=
\int
s_e(\vec{a},\lambda)\cdot s_e(\vec{b}, \lambda)
 [s_e(\vec{b},\lambda)\cdot s_e(\vec{c}, \lambda)-1]\, \rho (\lambda)\,d\lambda
\,.
$$
\end{widetext}
Since, under the stated hypothesis, the identity 
\[
\mathcal{P}(\vec{f},\vec{g})\equiv P(\vec{f},\vec{g})
\] 
follows (or so it seems if one's attention blinks as will be detailed), one can easily check that the \emph{Bell inequality} $(**)$ seems to be incompatible for some triples $(\vec{a},\vec{b},\vec{c})$ with the value $\mathcal{P}(\vec{a},\vec{b})
=-\vec{a}\cdot \vec{b}$ predicted by QM. Thus (failing to notice, or anyway ignoring the counterfactuals that are implicit in the process of deriving the inequality and comparing it to QM), Bell concludes:

\smallskip
\noindent \textit{``In a theory in which parameters are added to quantum mechanics to determine the results of individual measurements, without changing the statistical predictions, there must be a mechanism whereby the setting of one measuring device can influence the reading of another instrument, however remote."}

\smallskip
\noindent For more general inequalities that also now belong to the family of Bell's inequalities, see, \emph{e.g.,} \cite{CHSH}. For a lucid critique see \cite{PenaCettoBrody1972} (or p. 312 of \cite{Jammer1974} for a nice abstract of that paper).  See also Fine's book \cite{FineShaky} and references therein for related matters.
\begin{rem} [What can we learn from \cite{Bell}?] \label{rem:BadBoy}
The first sentence of Bell's paper \cite{Bell} reads:

\smallskip
\noindent \emph{``The paradox of Einstein, Podolsky and Rosen \cite{EPR} was advanced as an argument that quantum mechanics could not be a complete theory but should be supplemented by
additional variables."}

\smallskip
\noindent
The second sentence of \cite{Bell} is not much less irritating. It reads:

\smallskip
\emph{``These additional variables were to restore to the theory causality and locality."}

\smallskip
\noindent The fact that \cite{EPR} is only 4 pages long makes any such misquotation 
quite painful.  Moreover, since the HV provide (by the hypothesis in \cite{Bell}) the same statistics as QM, if QM is non-local, the HV theory is also non-local: no need for an inequality. Now if QM is local, the motivation for \cite{Bell} is quite questionable from the very point of view of the second statement in the introduction of \cite{Bell}: why try HV to restore locality to a local theory?  But since the inequalities are fully counterfactual, there is not much to extract from the analysis leading to the inequalities anyway, and of course as with any counterfactual, there is no way to experiment whatsoever (see also \cite{Tresser2005} where the counterfactual nature of the core of the problem becomes evident). So the celebrated 1964 paper by Bell \cite{Bell} appears as reducing to no physics at all or barely any. Much more will need to be told elsewhere, hopefully including a long overdue analysis of attacks by Bell on Jammer in \cite{Bell1976}.  Meanwhile I salute the provocative effect of \cite{Bell}, even if that small paper has caused so much confusion.
\end{rem}
\subsection{5.2: On the first part of Bell's 1964 paper}\label{sub:FirstPartBell}
Part 1 of \cite{Bell} has seldom been commented upon, with the noticeable exception of a paper \cite{deBroglie1974} by Louis de Broglie who refused to believe that
$\mathcal{P}(\vec{a},\vec{b})=-\vec{a}\cdot \vec{b}$ can occur ``for macroscopic separations",  the same words that Bell used in \cite{Bell1975}  where he made fun of de Broglie.  The truth is that the old master was wrong in his conclusions, but rightfully felt that something deserved attention there. What is implied by $\mathcal{P}(\vec{a},\vec{b})= -\vec{a}\cdot \vec{b}$ is indeed quite choking. Of course, QM \textit{does} tell us exactly that $\mathcal{P}(\vec{a},\vec{b})=-\vec{a}\cdot \vec{b}$, by combining WPR and the same considerations that lead to Malus Law for spin-$\frac{1}{2}$.  But the same Malus Law would, up to sign, give us the same correlation if a second Stern-Gerlach magnet was placed at the same angle, after the first magnet on the electron side. When the particles separate, they are mirror images of each other: 
\emph{all of a sudden, as the electron passes its first Stern-Gerlach magnet 
along $\vec{a}$, particle ${\bf p}$ which was mirror image of ${\bf e}$ \textbf{before} the magnet, starts 
to behave as if it would be the mirror image of the electron  \textbf{after} the magnet}. 

- On the one hand, one is tempted to say that this is such an obvious non-locality of QM that one should not wander that HV theories that imitate QM would also be non-local: see also Remark \ref{rem:BadBoy}.
Bell also used more trivial settings than EPR to show the non-locality of QM,
an example being, \emph{e.g.,} the analysis that  he provided 
on p. 89 of \cite{Bell7685}.  In that example (a single radioactive atom with detectors of $\alpha$ particles all around: if one detector detects, the probability for the others to
detect immediately falls to zero) the need for non-locality vanishes if, following EPR, we agree with (*), to the contrary of the belief that has entrapped Bell (and others in Quantum Physics: Bell wrote that he considered that most physicist do not take seriously enough that wave functions is all there is).

- On the other hand, I will provide evidence for the fact that all the phenomenology of EPRB, even in the Bell form, can be explained by recourse to conservation laws and to the Malus Law so that no appeal to any form of non-locality is necessary.  
%
\subsection{5.3: On the second part of Bell's 1964 paper}\label{sub:SecondPartBell}
The stress of Bell's historical paper \cite{Bell} and the fame of the subject rest on the inequality in Part 2, even if the inequality only deals with HV. The ``only" here is because the HV approach was already long abandoned at the time of EPR,  27 years before \cite{Bell}. This long abandon is true for physicists,
\emph{i.e.,} in particular not counting von Neumann, or rather counting him only as a Mathematician (despite his lasting contributions to physics):  von Neumann gave a false proof of no HV in his historical book \cite{vonNeumann} in 1932, thus about three years before \cite{EPR}. However Einstein and most other leaders of Physics had abandoned HV since 1927 or 1928 at least (about Einstein and HV, see \cite{FineShaky}, \cite{Jammer1974},  and \cite{Rosen1985}).  The hole in the 1932 proof by von Neumann was pointed out by Bell \cite{Bell1966} in a paper published after \cite{Bell} but apparently written before (see  \cite{Jammer1974}).  As Bell remarks in \cite{Bell1966}, the proof  by von Neuman was in fact moot by the time of \cite{Bell1966}, because of the formulations in 1927 by de Broglie and in 1952 by Bohm of their respective HV theories.  These theories are very similar to each other: for a full account by de Broglie on his 1927 HV approach, see his book published much later \cite{deBroglie19271956}, and for Bohm's work on HV see \cite{Bohm1952}.  The motivation of Bell (as he described it himself) in reviving HV was among other things to defend the de Broglie - Bohm HV theory as a pedagogical way into QM (at least). The exact technical statement of the result in \cite{Bell} was honestly characterized by Bell in that same paper as a result about HV, even if that was missed by many. More precisely, Bell's description of his own result is honest but wrong.  By the time when \cite{Bell} was published (29 years after EPR), counterfactual analysis was not yet at hand (although Einstein and others knew how to avoid them, in particular when they involved multiple measurements) and Bell missed the counterfactual character of the violation of his inequality (and later of generalizations thereof) by QM and by Nature, as described in \cite{PenaCettoBrody1972}.
%
%
\section{6: Bell's EPR by combining conservation laws and Malus Law \label{sec:EPR by conservation}}
\subsection{6.1: The third main EPR claim}
We are left with the formula
$\mathcal{P}(\vec{a},\vec{b})=-\vec{a}\cdot \vec{b}$ predicted 
by QM for the arbitrary angle (or \emph{Bell configuration}) of EPRB.  As long as we had $\vec{b}=\pm \vec{a}$, all was fine (if we forget EPR's concerns about completeness), in that sense that conservation laws could be invoked instead of the mysterious WPR and we had Claim \ref{claim:C0}.  The goal of the present section is to show that reduction to conservation laws is quite general by dealing with the wildest case, which is provided by Bell's version of EPRB. 
\begin{claim}\label{claim:C1'}
In the EPRB setting with measurements along vectors $\vec{a}$ for one particle and $\vec{b}$ for 
the other one, such that the angle $\theta$ from $\vec{a}$ to $\vec{b}$ is known but arbitrary, the fact that the correlation between the measurements, $\mathcal{P}(\vec{a},\vec{b})$, satisfies $\mathcal{P}(\vec{a},\vec{b})=-\cos(\theta )$ is the result of combining Malus Law with the conservation on angular momentum projected on any fixed axis.
\end{claim}

\noindent \textbf{Arguments for Claim \ref{claim:C1'}.} 
Now I make the measurement on ${\bf e}$ say, along $\vec{a}$ and the measure on ${\bf p}$
along $\vec{b}$.  Reasoning as in the arguments for Claim \ref{claim:C1},  and in particular using the instrumental realism from Claim \ref{claim:C0}, I get minus the measure on ${\bf p}$ along $\vec{b}$ on
(about) $LO$ for ${\bf e}$, and minus the measure on ${\bf e}$ along $\vec{a}$ on
(about) $OR$ for ${\bf p}$. Applying Malus Law on either side then yields the desired
result.  In view of the arguments for Claim \ref{claim:C1}, this concludes the arguments for Claim \ref{claim:C1'} 

\subsection{6.2: Back to weak realism}
\begin{rem}[Weak realism revisited]\label{rem:RealismRevisited}
After all the discussions that has been offered on instrumental realism, and having gone through the generality of the setting of Bell's version of EPRB, I feel it useful to review the meaning of weak realism. To be specific I will consider here the case of the spin of ${\bf p}$, but other observables would be amenable to parallel discussions. 

If $\vec{c}$ is any vector along which I want to test the positivity of the spin of ${\bf e}$, minus the spin of ${\bf e}$ will be the spin of ${\bf p}$ before I read it by the experiment if I am in a case when realism holds true. If ${\bf p}$ is just any particle, I can choose any \textbf{one} such $\vec{c}$ to extract this knowledge, and that may be the extent of knowledge that I get. With realsim, this knowledge corresponds to ${\bf p}$ from before the measurement and for the future as long as no event happens to ${\bf p}$ in any of the directions of time. With no realism, only the future knowledge is accessible but for (at least most) practical purposes, this is the only direction of time that matters.

If now ${\bf p}$ is one member of an EPR entanglement:

\noindent
- I know that I have realism in the form of instrumental realism, 

\noindent
- I can choose two directions along which the value of the spin of ${\bf p}$ can be extracted,

\noindent
- The reading from the direction chosen to measure the spin of ${\bf e}$ being taken at $t_0$ (Lorentz invariantly) before measuring the spin of ${\bf p}$ along some vector at time $t_1$, the values of the spins of ${\bf p}$ along all vectors may make sense (and for coherence I think, has to make sense) at $t_1'$ just before $t_1$ (but Lorentz invariantly after $t_0$), but,

\noindent
- The only values that I can extract at $t_1'$ are those along the two vectors chosen at $t_0$ for 
${\bf e}$ and at $t_1$ for ${\bf p}$, and this knowledge of what holds true at $t_1'$ will be obtained not earlier than $t_1$ in the referential  where the second measure is performed.

\noindent
- From $t_1$ on, the knowledge of only one of the future spins of ${\bf p}$ will be accessible: the spin along the vector chosen to measure the spin of ${\bf p}$ at $t_1$ since, 

\noindent
- As usual, any other one is eradicated by the measurement process, the same process that forces double knowledge to be only possible in $(t_0, t_1)$.
\end{rem}
%
%
\section{7: On the decay of geometry at small enough scale\label{sec:Decay}}
The simple classification of counterfactuals ideas presented in
Section 2 have allowed me to recast in the ``suspicion against counterfactuals
mode" some observations that are rather simple but have
far reaching consequences, as I try to expose next.

\subsection{7.1: Scale-based counterfactuals}
As explained in layman terms by Einstein in
\cite{Einstein}, \emph{length and time measurement tools are
essential foundations to mechanics,} a profound statement, despite
the fact that it can be very plainly stated. This leads me to
propose the following :

\smallskip
\noindent {\bf Fundamental observation.} \emph{There is no mechanics, without description and measurement of trajectories, so that, as soon as the atomic nature of the world is accepted, very 
small length measure tools and clocks just do not exist, depriving \emph{space}, \emph{time}, 
and \emph{causality} (and other concepts beside those like \emph{motion} that depend on these 
three ones) of any physical sense at a small enough scale. Furthermore, while the size of 
small-enough is ill-defined, and probably not universal anyhow, one can expect that this is much 
bigger than the Plank scale.}

\smallskip
I have been obsessed by the non-feasibility of tiny clocks since I was fourteen, 
but could never use that remark by itself. It is while reading Einstein again, and getting caught by his emphasis on the need for measure instruments to build mechanics, that I realized (without having yet recognized myself the danger of counterfactuals that were already well known to many others) that it was a counterfactual to try measuring space-time at small enough scale, small enough for 
no measurement to be fundamentally doable according to all one knows of Physics. The consequence as expressed by the above simple but fundamental observation is then clear.  This implies the following:

\begin{claim}\label{claim:C2}
Geometry does not have any physical sense at small enough scale.
\end{claim}

\noindent
This claim admits in turn the following immediate corollary:

\begin{cor}\label{cor:CC2}
Causality does not make any physical sense at such scales.
\end{cor}
\begin{rem}[A Copenhagen Interpretation view]
Notice that thanks to the basic point of view presented in Claim \ref{claim:C2}, the need for Classical Mechanics 
to even formulate Quantum Mechanics turns from mysterious and annoying to rather natural: it plausibly becomes a \emph{need} for any Physics of the very small.
\end{rem}

\begin{rem}[Realism and Pauli's angels]\label{rem:angels}
Realism takes a new meaning in view of Claim \ref{claim:C2}: the ``coordinates" that make sense for the microscopic world, or rather what replace coordinates, are different from the way we express observables by measurements made by us as macroscopic entities utilizing (macroscopic) instruments to access information about the microscopic world. The fact that two conjugate quantities cannot coexist in general does not imply the doom of realism. It ``only" means that something is lost in the measurement process, so that realism cannot happen on conjugate variables on a given particle, except, I claim with Schr{\"o}dinger (see my Claim \ref{claim:C1}) for special cases like particles that participate to EPR type entanglements.  In general, realism may well always be true ``for the angels" (borrowing there the angels from Pauli's letters to Born reprinted in \cite{BornEinstein} about Born dispute with Einstein) with no mean that I know of to check if it holds or not.  Does this have anything to do with Einstein's call to get out of the usual coordinates in order to go beyond QM? Who might know?\end{rem}

A question that remains is whether there is some understandable structure, such as topology or combinatorics, that makes \emph{physical} sense at  arbitrary small scale. For instance does any structure make sense ``in any neighborhood of any point" if topology makes sense, given that according to Claim \ref{claim:C2}, there is no more physically meaningful size at such scale? Being able to give a physical sense to the words  ``in any neighborhood of any point" would be a crucial test.
One should also take care that once there are no more scales, ``small" and ``big" cannot any more be that different from each other. Indirect measures such as using mass to infer sizes may be quite deceptive.  When one gets to large enough scales, as measured from our scale, the relative sizes 
of where geometry decays becomes negligible, so that geometry becomes approximately valid. At intermediate scales, before geometry makes approximate sense, weaker structures such as topology and order structure may well get physical meaning.  All these things deserve serious investigation.

Because of the decay of geometry at small scale, the geometrizable part of the Universe appears as a sort of sponge with fuzzy holes filled with a-geometric parts all around (what one could call \emph{``the Space-Time Fuzzy Sponge"}).  
The holes in the sponge, or for those who do not like the sponge image, the
a-geometric components of the Universe, are not expected to be
quantized: the number of quantal players needed to come close to
the classical regime seems to not be universal as quantum
super-fluids  and so called SQUIDs (Superconducting QUantum
Interference Devices) indicate (for an argument in the opposite
direction, see \textit{e.g.,} \cite{Omnes}): the fact is that non
interacting quantum components can build very large quantum
objects whose geometry makes sense, but at the macroscopic level.

\smallskip
Many statements by Heisenberg (see \cite{Heisenberg}) are quite compatible with Corollary \ref{cor:CC2}.  However, Heisenberg comes short in \cite{Heisenberg} of formulation that statement or a similar one.  What seems really new (and the only thing most probably according to Arthur Fine) is the counterfactual analysis approach to Corollary \ref{cor:CC2}.  This approach reinforces the similarity of the loss of physical sense of geometry at small enough scale with the fact that most natural numbers make no physical sense.

\smallskip
The EPR-related part of the present paper aims at helping kill the legend of non-locality of QM associated to EPR (others have already began that process, but the spread of the disease is huge). On the other hand (and this transpires in the way so many people were ready to accept 
the EPR-motivated strong non-locality statements), there is (or seems to be) some \emph{sort of (weak) non-locality} of QM associated for instance to simple double slits settings, wave mechanics and path integrals \cite{FeynmanHibbs}.  
The decay of geometry at small enough scale may help us understanding this still quite mysterious (weak) non-locality. However, the best point might be that comprehension of QM becomes an acceptably impossible task.
\begin{rem}[Counting non-locality properties]\label{rem:Vaidman}  
To the direct contrary of the point of view that I defend, some consider in fact that 
\emph{the (strong) non-locality of QM associated to EPR} and \emph{the (weak) non-locality of QM as in double slits settings} are the same and that the strong-weak dichotomy that I have used here 
is meaningless. From what I understand from the conversation that I had the chance of having with him, it seems to me that Lev Vaidman is in that camp (of course, this did not prevent the discussion from being very useful to me, with lots of points of agreement and many of disagreement: see also \cite{Tresser2005}).
\end{rem}
\subsection{7.2: Experimental test.}
Reconsider the experimental setting for diffraction and interferences with neutrons (see for example the review in \cite{Neutrons}) or other particles. Performing the experiment with heavier and heavier nuclei, and then if needed more complex objects, one can expect to see less and less interference as the approach to classicality builds up geometrization and predictability of the trajectories according to the views defended on the basis of the decay of geometry at small enough scale.  However, geometrization could happen at once at some threshold value, possibly then with hysteresis. I remark that the experiment that I propose would test
likely consequences of the decay of geometry at small scale, rather than the decay itself or logically provable consequences of it (but one should keep in mind that the the decay of geometry may, but does not have to, mean the decay of mathematization altogether).
\subsection{7.3: Introducing wave pseudo-functions.}
The quantum descriptors such as fields of wave functions should
depend on \emph{``about $x$ and about $t$"} rather than on $x$ and
$t$, al the proper formalism to express that eludes us at
this (approximate) time. The truth is that \emph{pseudo wave functions}, or
rather \emph{wave pseudo-functions}, appear to often be
well approximated by usual wave functions, since parts of Quantum
Mechanics and Quantum Field Theories work quite fine. 

Other forms of time destructions and geometry changes or geometry weakening at about the Planck scale, under extreme conditions or not, have appeared following ideas initiated by de Witt and Wheeler  
\cite{DeWitt}, \cite{Wheeler}: see in particular \cite{Rovelli} and \cite{HellerSasin} which builds on non-commutative geometry \cite{Connes}: when these approaches come closest to the simple one proposed here, they are epistemologically most different from it, but the reader is entitled to her or his own opinion on this matter.
\begin{acknowledgments}

\medskip
\textbf{Acknowledgments.}  Ed Spiegel has been the main
witness and support of my internal fights, while Michel le Bellac, Pierre Hohenberg, Larry Horwitz, Jenann Ismael, Marco Martens,  Itamar Pitowsky, Itamar Procaccia, Oded Regev, Nils Tongring,
and Lev Vaidman also gave me substantial listening time, some of them  when I was saying more non-sense than else.  Seminars delivered at NYU and the Technion were also quite precious: I am indebted to Tycho Sleator and Yosi Avron for enduring my requests before enduring the talks. Both audiences helped me a lot refine some concepts and arguments.  Also listening to me and bombarded by mails were Pierre Coullet, Jean Marc Gambaudo, Albert Libchaber, John Milnor, Dan Rockmore and Dennis Sullivan. Chiara Toniolo's support was also of great value. Lately, but no least, I got some precious remarks from Arthur Fine. It was also useful to test some of my explanations on my very dear Bernard Laleuf and David Serre, two \emph{civilians}, \emph{i.e.,} 21$^{\rm st}$ century men interested in Physics.  At last I could not have written this paper, especially as sick as I am now, without the support
of my lovely wife L{\'e}a and my adorable children Yuval, Rachel, Yga{\"e}l, and Liza and so 
many sisters, brothers and other friends, colleagues and doctors and nurses who have so much
contributed to save my life, and still help me keep my pain under control.
\end{acknowledgments}

\bibliography{basename of .bib file}

\end{document}